\documentclass[conference,a4]{IEEEtran}
\usepackage{amsmath,amsfonts}
\usepackage{algorithmic}
\usepackage{algorithm}
\usepackage{array}
\usepackage[caption=false,font=normalsize,labelfont=sf,textfont=sf]{subfig}
\usepackage{textcomp}
\usepackage{stfloats}
\usepackage{url}
\usepackage{verbatim}
\usepackage{graphicx}
\usepackage{cite}

\newcommand{\mynotex}[1]{}

\IEEEoverridecommandlockouts
\usepackage{graphicx} 
\usepackage{balance}

\title{Incentivising green video streaming through a 2-tier subscription model with carbon-aware rewards}

\author{\IEEEauthorblockN{Vasilios A. Siris, Adamantia Stamou, George D. Stamoulis, Konstantinos Varsos}
\IEEEauthorblockA{\textit{Department of Informatics, School of Information Sciences and Technology,} \\
\textit{Athens University of Economics and Business, Greece}\\
\{vsiris, stamouad, gstamoul, kvarsos\}@aueb.gr}}

\begin{document}

\maketitle
\thispagestyle{plain}
\pagestyle{plain}

\begin{abstract} 
We investigate incentives for reducing the carbon emissions of video streaming that depend on the energy consumption of segments in the end-to-end video delivery path, the carbon intensity, and the user type, i.e., quality-sensitive and green or environmentally conscious users.
The incentives can be offered through a practical 2-tier  subscription model with a discount and carbon rewards, which gives providers the flexibility to reduce the quality for up to a maximum percentage of videos within a time period, such as one month.
The key features of our approach are i) it is preferable to offer subscriptions where the reduced-quality tier is set one resolution level below the  resolution required for maximum user satisfaction; ii) when a video is streamed from a local data center, the maximum percentage of videos streamed at a lower quality depends solely on the carbon intensity  and the average intensity cap, whereas the incentives also depend on the users' level of environmental consciousness;  iii) when a video can be streamed from a local or a remote data center with  different carbon intensities,  the maximum percentage of videos streamed at  lower quality and the incentives depend on the relative carbon intensity and energy consumption at the data centers, and the additional network energy costs  from the remote data center.  
\end{abstract}

\section{Introduction}

On-demand streaming constitutes 54\% of the total downstream volume in fixed networks and 57\% in mobile networks, while live streaming is another 14\% and 7\% of the total downstream volume for fixed and mobile networks, respectively~\footnote{The Global Internet Phenomena Report, March 2024, https://www.applogicnetworks.com/phenomena }.
Hence, video streaming contributes significantly to energy consumption and carbon emissions due to data transmission of online services. Indeed, it can be argued that video streaming has become the ``killer-app'' at the household level~\cite{Mad++23}.
At the same time, the huge increase in demand is pushing network capacity to its limits, making it increasingly important for providers to manage network resources ensuring that the offered Quality of Experience (QoE) meets customer expectations.
The need for efficient management of network resources, along with the rising concerns for sustainability and carbon emissions produced by  ICT infrastructure and services, creates the need to investigate the tradeoffs of energy efficiency, carbon emissions, and user QoE. 

Studies have highlighted the impact that  user behavior, combined with factors from the provider side,  have on the carbon emissions of video streaming~\cite{Sus++20,Seg++23}. Moreover, incentive-based policies for reduced video quality 
can influence future energy consumption and carbon emissions~\cite{Mad++23}. This last study shows that the impact of such policies depends on the number of subscribers and their behavior, the energy efficiency of end-user devices, the energy consumption of video data transfer, which includes the transmission network and the data center, and the use of low carbon intensity energy sources. Users can stream video over different access technologies, with widely varying energy consumption, while the data centers hosting  videos can also have different energy consumption as well as varying carbon intensities.
To be effective, incentives offered to users by video providers and network operators  should consider all these factors. 

This paper contributes to the design and investigation of practical incentive schemes where users give providers the flexibility to reduce the quality of the video they serve during periods of high carbon intensity, in exchange for reduced subscription costs and carbon-aware  rewards.
Specifically, our contributions are the following:
\begin{itemize}
    \item We present a framework for determining the user's utility loss when a video is streamed at reduced quality; for users to accept the reduced video quality, this utility loss must be compensated through reduced  subscription costs and carbon-aware rewards.
    \item We investigate the impact that the energy consumption of different segments along the end-to-end path from a data center to the end user can have on the design of incentives.
    \item Utilizing  empirical measurements of the daily carbon intensity, we present a procedure for determining the incentives that a provider should offer to achieve  a target average carbon intensity cap, both for cases in which a video is streamed from a local data center and when the video is available both at a local and a remote data  center.
    \item We present a practical  two-tier subscription model that gives providers the flexibility to reduce the quality for up to a specified maximum percentage of videos during a subscription period, e.g., one month, based on the actual carbon intensity, in exchange for offering users a reduced subscription cost and carbon-aware rewards.
\end{itemize}
The remainder of the paper is structured as follows: In Section~\ref{sec:related_work} we discuss related work, identifying the contribution of this paper. In Section~\ref{sec:utility_incentives} we discuss the user utility for video streaming and the incentives based on it. In Section~\ref{sec:energy_consumption_e2e} we use energy consumption data of segments on the end-to-end video delivery path to discuss the implications for the design of incentives. In Section~\ref{sec:carbon_aware_dc_selection} we present a simple model for carbon-aware data center selection and in Section~\ref{sec:incentives} we discuss the energy reduction and incentives for different user types, for the case of a local data center and for the case of a local and remote data center.  Finally, in Section~\ref{sec:2-tier} we present a practical 2-tier subscription model with carbon-aware rewards based on the incentive model analyzed in the previous sections. Finally, Section~\ref{sec:conclusions} concludes the paper identifying directions for further investigation.

\section{Related work}
\label{sec:related_work}

\mynotex{
\begin{itemize}
\end{itemize}
}

The analysis in~\cite{Mad++23} shows that energy consumption for data transmission is significant compared to the energy consumption of end-user devices. Indeed, this holds for all three scenarios investigated, each with different assumptions for  the energy efficiency of end-user devices and data transmission, namely a baseline scenario where current energy efficiency trends continue, a gray scenario that corresponds to low energy efficiency improvements, and a green scenario that corresponds to higher energy efficiency levels.
Furthermore, the study quantifies the impact of energy consumption at the data center, identifying that in all three of the aforementioned scenarios, the relative share of data centers’ electricity consumption increases, which supports the data center's increasing importance for reducing  the energy consumption of future video streaming services.
Overall, this study identifies that energy consumption depends on four key parameters: the number 
of subscribers, energy efficiency of end-user devices, energy consumption of data transmission, and video resolution, while carbon emissions depend on these factors along with the carbon intensity of the energy sources. The paper suggests that video streaming at lower resolution can be triggered by regulatory intervention, through enforcement or incentive-based policies, without however proposing specific mechanisms.

The work in~\cite{Sus++20} investigates the impact of user behavior on carbon emissions, highlighting key factors being the 
choice of viewing device and video viewing patterns. Moreover, this work found that a very small percentage of users modify the default video resolution settings, suggesting that a lower default resolution can be set by the provider. 
The investigation in~\cite{Seg++23} concludes that
simply informing users about the environmental impact of video streaming can influence their behavior, which supports the importance of increasing awareness of the environmental impact of video streaming services. 
However, they also found that introducing specific carbon reduction targets did not further reduce  carbon emissions. 
Possible reasons are that the target carbon emissions reduction was not sufficiently difficult or that all of the achievable carbon gains were already achieved with the  increased awareness. 
This result suggests that  targeted feedback is necessary, such as in the form of monetary or reward incentives, which is the focus of this paper. Moreover, the observation also suggests that incentives should be provided when  carbon emissions reduction is actually needed, which we consider by taking into account actual carbon intensity measurements.

The work in \cite{Hos++23} quantifies the trade-off between the QoE of video streaming services and CO2 emissions. 
User satisfaction is represented by the Mean Opinion Score (MOS), which is a logarithmic function of the video bitrate. The model includes a greenness factor, which depicts the user's  willingness to accept a reduced  MOS  for a higher carbon emissions reduction. 
Kleinrock’s power metric, defined as the ratio of the QoE to the corresponding amount of carbon emissions, is used to determine optimal video bitrates that  balance QoE and emissions. 
The work in \cite{Bin++23} also investigates the trade-off between sustainability and QoE for video streaming, while also considering the energy consumption of the devices where the video was viewed. Additionally, this work conducted a crowdsourcing survey that showed that users were willing to incur a reduction of the video quality to the next lower quality level, but were  reluctant to reduce the video quality by multiple quality levels, despite such a reduction providing higher energy consumption gains. Investigation of the device energy consumption and video QoE tradeoff is also the focus of the work in \cite{Bin24}.
We consider the same utility function for video streaming investigated in \cite{Hos++23,Bin24} to design incentives and make the following unique contributions: First, we investigate how the energy consumption of all segments in the end-to-end video delivery path, including multiple  access technologies and CDN sizes, affects the design of the incentives. Second, 
we propose a 2-tier subscription model for exposing the incentives to users. The model gives providers the flexibility to reduce the video quality when the carbon intensity is high and in exchange the users selecting such a subscription receive a discount and carbon rewards.

Recent literature has expressed concerns on overestimating the energy consumption in different segments of the end-to-end path, including the mobile and fixed access network and the data center~\cite{Kam20}.
Energy consumption overestimation can result in  top-down approaches that estimate the energy consumption per data unit transferred (kWh per GB) considering aggregate power consumption numbers, such as the total global energy consumption of data centers. Such an approach fails to separate the idle energy consumption of devices from the incremental or traffic-dependent energy consumption~\cite{Myt++24}.
For this reason,  in
Section\ref{sec:energy_consumption_e2e} we consider energy consumption data obtained exclusively using bottom-up power models and measurements for segments or equipment in the end-to-end video delivery path. Moreover, we consider solely the incremental energy consumption of a video stream, expressed in Watts per Mbps.
The justification for considering only the incremental energy consumption and not the idle energy consumption is because the latter is not affected by the video bitrate, hence when the video quality is reduced. Incentives, which are the focus of this paper, influence the video quality and bitrate, which in turn affects the energy consumption only through the incremental energy consumption component.

\section{User utility for video streaming and incentives}
\label{sec:utility_incentives}

\mynotex{
\begin{itemize}
\item For mobile devices, which have a smaller screen than tvs and laptops, it can make sense to consider that the maximum video quality which achieves utility 1 is FHD. Cite reference Sustainability vs. Quality of Experience: Striking the Right Balance for Video Streaming
https://dl.acm.org/doi/fullHtml/10.1145/3708973.3708975

\item for MOS use term logistic regression model
\end{itemize}
}

\begin{figure}[tb]
    \centering
    \includegraphics[width=0.95\linewidth]{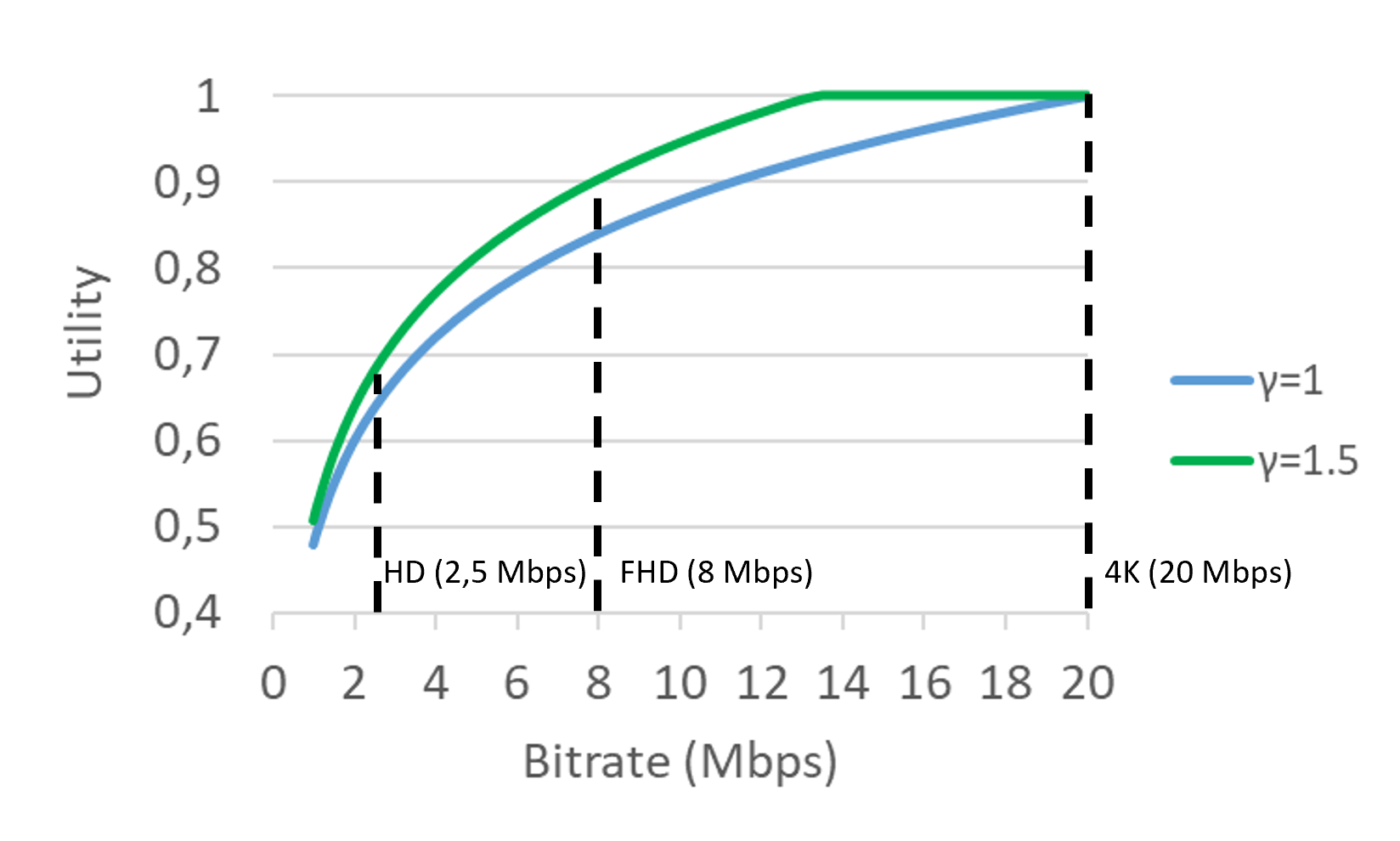}
    \vspace{-0.2in}
    \caption{User utility for video streaming  when the maximum bitrate is $x_{\max}=20$ Mbps (4K resolution). A green user ($\gamma=1.5$) achieves the same utility as a high-quality user ($\gamma=1$), but for a lower bitrate.  
    }
    \vspace{-0.2in}
    \label{fig:utility}
\end{figure}

We consider a  user utility for video streaming based on the Mean Opinion Score (MOS) formulation in \cite{Hos++23}:
\begin{equation}
\resizebox{0.99\hsize}{!}{%
$MOS_{\gamma}(x) = \frac{4}{\log (x'_{\max}) - \log(x_{\min})}\log (x) + \frac{\log (x'_{\max}) - 5 \log (x_{\min})}{\log (x'_{\max}) - \log (x_{\min})}$ \, ,
}
\label{eq:MOS}
\end{equation}
where $x_{\min}$ is the minimum bitrate (set to 0.2~Mbps),
$x'_{\max}=x_{\max}/\gamma$, with $x_{\max}$  the highest bitrate at which MOS=5, and $\gamma$  a ``greenness factor''. A high-quality (HQ) user has $\gamma=1$ and achieves MOS=5 for bitrate equal to $x_{\max}$, while a green user with greenness factor $\gamma>1$ achieves MOS=5 for a lower bitrate $x'_{\max}=x_{\max}/\gamma < x_{\max}$. The MOS is normalized to give a user utility $U_{\gamma}(x)$ in the range $[0,1]$:
\begin{equation}\label{eq:utility}
U_{\gamma}(x)= \frac{1}{5} MOS_{\gamma}(x) \, .
\end{equation}
Figure~\ref{fig:utility} shows the utility for a high-quality user ($\gamma=1$) and a green or  environmentally conscious user ($\gamma=1.5$), when $x_{\max}=20$ Mbps, which corresponds to 4K video resolution. Observe that the green user has  a higher valuation for the same bitrate compared to a high-quality user, when the latter's utility is smaller than 1. Alternatively, a green user achieves the same utility as a high-quality user, but for a lower bitrate. Moreover, observe that for lower utility values, the bitrates at which high-quality and green users achieve the same utility converge. As we discuss below,  this influences the video quality reductions that a provider should target with incentives.
In the experiments reported in later sections, we assume that users can reduce their video quality from 4K to FHD (60~Hz), which corresponds to a bitrate of 8~Mbps and HD (720p), which corresponds to  bitrate of 2.5~Mbps.
The utility loss when users reduce their video quality from 4K to FHD and HD is $\Delta U^{\text{4K}\rightarrow\text{FHD}}_{\gamma} =  U_{\gamma}(x_{\text{4K}}) -  U_{\gamma}(x_{\text{FHD}})$ 
and $\Delta U^{\text{4K}\rightarrow\text{HD}}_{\gamma} =  U_{\gamma}(x_{\text{4K}}) -  U_{\gamma}(x_{\text{HD}}$), respectively, where $x_{\text{4K} }=20$~Mbps, $x_{\text{FHD} }=8$~Mbps, and $x_{\text{HD} }=2.5$~Mbps; the utility loss for reduced video quality is shown in Table~\ref{tab:utility_loss}.
For smartphones, which have a smaller screen size than TVs and desktops/laptops, if we assume that video quality FHD achieves utility equal to one, then we can set $x_{\max}= x_{\text{FHD}}$ in (\ref{eq:MOS}).

We make two important observations from Table~\ref{tab:utility_loss}.
\begin{itemize}
\item Reducing the video quality from 4K to FHD reduces the bitrate by 60\% for both the high-quality and green users, while  the utility reduction is only 16\%  for high-quality users and 10\% for green users. A reduction from 4K to HD reduces the bitrate by 87.5\% for both user types, while  the utility is reduced 36\% for high-quality users and 32\% for green user.
\item Reducing the video quality from 4K to FHD results in utility loss for an HQ user that is 60\% higher than the utility loss for a green user, whereas reducing the quality from 4K to HD, i.e., a larger video quality reduction, results in a utility loss for a high-quality user that is only 12.5\% higher than that of a green user.
\end{itemize}
The first observation shows that a smaller reduction of video quality, i.e., from 4K to FHD rather than to HD, provides a  throughput reduction that is relatively higher than the corresponding utility reduction. As we  discuss in Section~\ref{sec:incentives}, this makes it preferable for a provider to offer incentives that involve  a smaller reduction of video quality, even if that requires reducing the  quality of more videos within  a month to achieve an average carbon intensity target.

The second observation shows that the behavior of a green user and a high-quality user differ more for a small reduction of the video quality. On the other hand, for a high reduction of the video quality their behavior is closer, since their utility loss differs less. This further supports the  above remark that smaller reductions in video quality are preferable for providers, allowing them to benefit more from the green user behavior and educating users to be environmentally conscious.

\begin{table}[tb]
\begin{center}
\caption{Utility loss for high-quality and green users, when they reduce their video quality from 4K to FHD and HD.}
\label{tab:utility_loss}
\begin{tabular}{|l|c|c|}
 \hline User type  & 4K $\rightarrow$ FHD &  4K $\rightarrow$ HD   \\ \hline \hline
high-quality & 0.16 & 0.36    \\ \hline
green & 0.10 & 0.32 \\ \hline
\end{tabular}
\end{center}
\vspace{-0.2in}
\end{table}



\mynotex{
\begin{itemize}

\end{itemize}
}

To incentivise a  user to reduce their video quality, the provider should offer incentives that compensate the user's utility loss. Moreover, when users reduce their bitrate, they reduce their energy consumption, which in turn reduces the provider's usage costs. The reduced usage costs can be passed to the users in the form of a monetary discount $s_{\delta \text{quality}}$.
Hence, the net utility loss $NL_{\gamma,\delta \text{quality}}$ can be written as 
\begin{equation}
NL^{\delta \text{quality}}_{\gamma} = \Delta U^{\delta \text{quality}}_{\gamma}  -s_{\delta \text{quality}} \, .
\label{eq:net_utility_loss}
\end{equation}
As shown in Table~\ref{tab:utility_loss}, the utility loss for  high-quality users is higher than that of  green users. Hence, the incentives that need to be offered to the green users are lower than those that are necessary for the quality-sensitive users, highlighting the importance of educating users on environmental consciousness.
However, the monetary discount $s_{\delta \text{quality}}$ depends solely on the reduction of video quality, i.e., it does not depend on the user type. 
The net utility loss can be provided to users in the form of rewards or as an additional monetary discount on top of $s_{\delta \text{quality}}$. We discuss this further in Section~\ref{sec:incentives} where we investigate incentives based on empirical measurements of the carbon intensity and in Section~\ref{sec:2-tier} where we discuss the proposed 2-tier subscription model.
%


\section{Energy consumption in the end-to-end path}
\label{sec:energy_consumption_e2e}

\mynotex{
\begin{itemize}
\item Decide if the estimations  based on the power measurements in the Chinese base station devices will be included.
\end{itemize}
}

The incremental energy consumption for different segments along the end-to-end path of a video stream is shown in Table~\ref{tab:energy_consumption}. This table shows only the incremental or per video traffic volume energy consumption. In addition to the usage-based energy consumption, the total energy consumption includes a static or traffic-independent component consumed when devices do not process traffic (i.e., when devices  are idle); however, the focus of this paper is incentives, which influence only the traffic-dependent (dynamic) part of energy consumption, for this reason, we only  consider the incremental energy consumption. Nevertheless, although the incremental energy consumption influences the design of incentives, the static energy consumption component influences the impact of incentives.
Also, the static energy consumption affects the overall discount of a  subscription involving reduced quality streaming.
We consider the energy consumption data in Table~\ref{tab:energy_consumption} to discuss their implications to the incentives; however, our approach for designing incentives is independent of specific measurements for the energy consumption in different segments along the end-to-end video path.

\begin{table}[tb]
\begin{center}
\vspace{+0.05in}
\caption{Incremental energy consumption of different segments in an e2e path.}
\label{tab:energy_consumption}
\begin{tabular}{|l|l|r|}
 \hline Segment  & Energy consumption  & \footnotesize{Ratio to wired}   \\
  & (W/Mbps) & \footnotesize{access+core}    \\
\hline \hline
Core network & 0.03 & 0.60 \\ \hline
Wired access & 0.02 & 0.40 \\ \hline
Wired access + core & 0.05 & 1.00 \\ \hline
4G access$^{\text{a}}$ & 1.5 (suburban) & 30.00 \\ \hline
4G access$^{\text{b}}$ & 8.86 (dense urban) & 177.20 \\ 
 & 14.9 (wilderness-to-urban) & 298.00 \\ \hline
5G access$^{\text{b}}$ & 4.2 (dense urban) & 84.00 \\ 
 & 6.3 (wilderness-to-urban) & 126.00 \\ \hline
6G access & 0.42 (10x lower than 5G) & 8.40 \\ \hline
Data Center & 0.01 (very large CDN) & 0.20 \\ 
 & 0.025 (large CDN) & 0.50 \\ 
 & 0.05 (medium CDN) & 1.00 \\ 
 & 0.09 (small CDN) & 1.80 \\ \hline
\end{tabular}
\end{center}
\vspace{-0.2in}
\end{table}

The data in Table~\ref{tab:energy_consumption} for the core network and for wired broadband access are obtained from \cite{Mal20}; see also \cite{Myt++24}.
The energy consumption data on line ``4G access$^{\text{a}}$'' are also  from~\cite{Mal20} and consider energy consumption measurements from a 4G base station in a suburban environment, including cooling, power conversion,  baseband and data transmission. 
The results for ``4G access$^{\text{b}}$'' and ``5G access$^{\text{b}}$'' are obtained from \cite{Gol++23}, and include the energy consumption of active cooling, main supply, and power converters. The data for 4G base stations in~\cite{Gol++23} consider linear regression of actual power and data traffic measurements at the same base station. The results  show that the incremental energy consumption has a strong linear relationship with the data volume, which is confirmed by other literature studies; see~\cite{Gol++23} and the references therein. Note that the linear model considers  measurements from multiple devices connected to the same base station. The energy consumption for an individual mobile device depends on the propagation characteristics from the base station to that device.
The energy consumption for 5G base stations is obtained by scaling the 4G model with  respect to bandwidth, number of data streams, and technological improvements. 



The results in Table~\ref{tab:energy_consumption} show a wide variation of  energy consumption of 4G base stations:  ``4G accesss$^{\text{a}}$'' from~\cite{Mal20} and ``4G access$^{\text{b}}$'' from~\cite{Gol++23} differ by a factor of six. In general, the energy consumption of 4G/5G base stations depends on their configuration, including the duplexing scheme, bandwidth, MIMO configuration, deployment location, etc. 
6G technology is expected to be 10 times \cite{Vis++25} to 100 times \cite{Ahm++25} more energy efficient than 5G. With a 100x improvement the incremental energy consumption of 6G is the same order of magnitude as the wired access and core network energy consumption shown in Table~\ref{tab:energy_consumption}.

The methodology presented in this paper for determining incentives and designing subscription models to provide these incentives is independent of the actual energy consumption values, which nevertheless  will affect the achievable energy and carbon emission savings. Moreover, the relative energy consumption of different segments in the end-to-end video delivery path can indicate that some segments have a larger contribution to energy and carbon emission savings. 
Moreover, 4G, 5G, and 6G technologies will coexist in future mobile networks, hence the energy consumption of  cellular access can vary significantly; this motivates the need to perform measurements along the end-to-end delivery path  to assess the impact of the segments in different deployments and effectively design incentive schemes.

The energy consumption for data centers is obtained based on the bottom up power model from \cite{Gue++23}, which considers best practices for CDN dimensioning by Netflix.
To consider only the incremental energy consumption, we assume that the part of the total energy consumption due to usage for video flash servers is 65\%, which follows the observations from \cite{She++24}, whereas for edge routers we consider the percentage to be 50\%. Indeed,   \cite{She++24} indicates that the percentage of total power  consumed during server idle periods is steadily decreasing, and is expected to reach 27\% by 2028; this suggests that the dynamic or usage-based component of energy consumption  in data centers will become increasingly important. 
Based on the above, we obtain the incremental energy consumption shown in Table~\ref{tab:energy_consumption} labeled "very large CDN".
Moreover, we  adapt the model to consider flash servers and edge routers for smaller size CDN deployments, including large, medium, and small, whose energy consumption with  respect to the energy consumption  of fixed broadband access (wired access +core network) are shown in  Table~\ref{tab:energy_consumption}. The range of CDN sizes  will allow us to quantify the impact of data center energy consumption for video streaming services in the case of fixed broadband access.

\subsection{Cellular access}


\begin{figure}[tb]
\begin{center}
\begin{tabular}{c}
\begin{minipage}[b]{1\linewidth}
\centering
\hspace{-0.28in}
\includegraphics[width=2.9in] {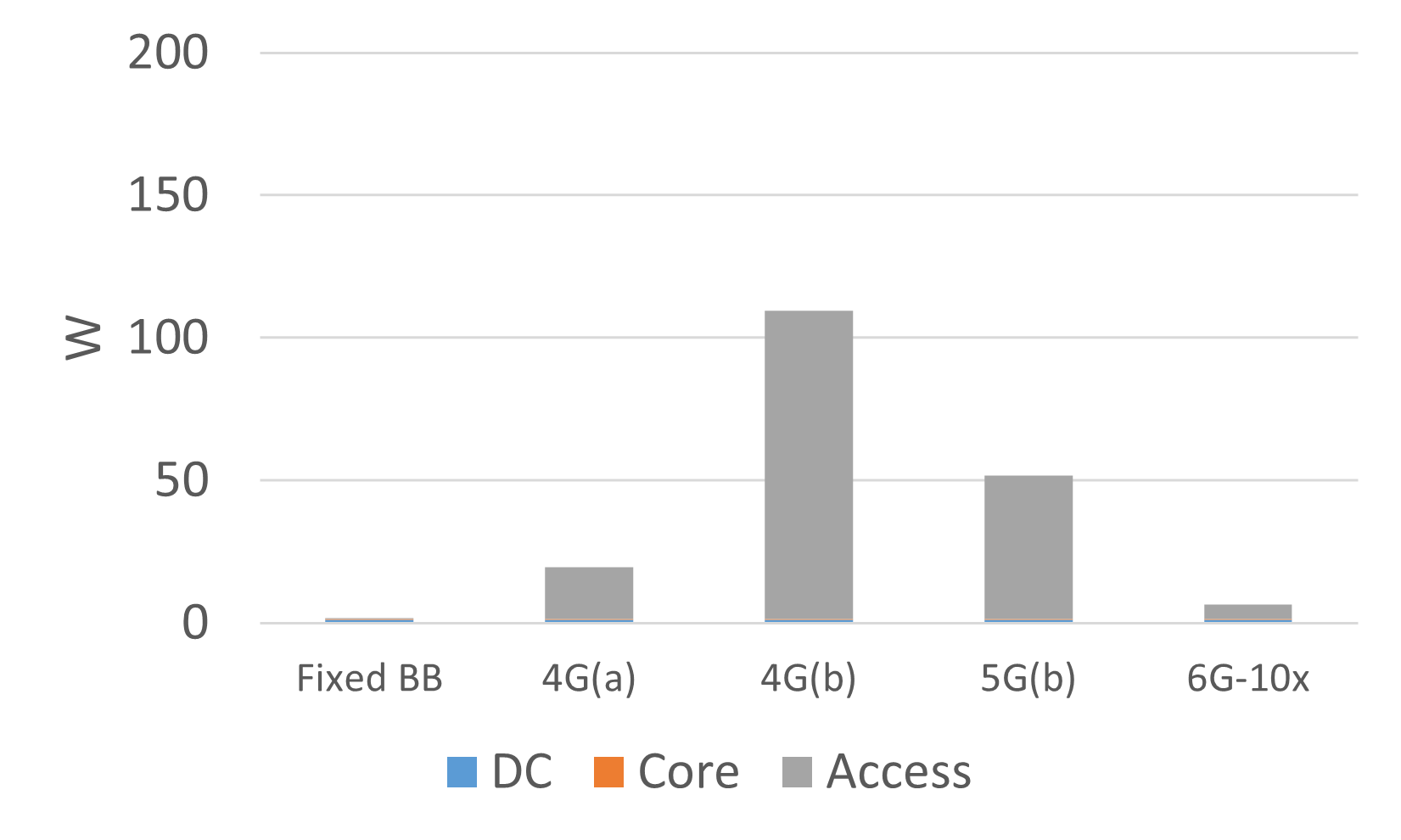}\\
{\footnotesize \hspace{-0.22in}\small{(a)  4K to FHD}}
\end{minipage} \\ \\
\begin{minipage}[b]{1\linewidth}
\centering
\hspace{-0.28in}
\includegraphics[width=2.9in]{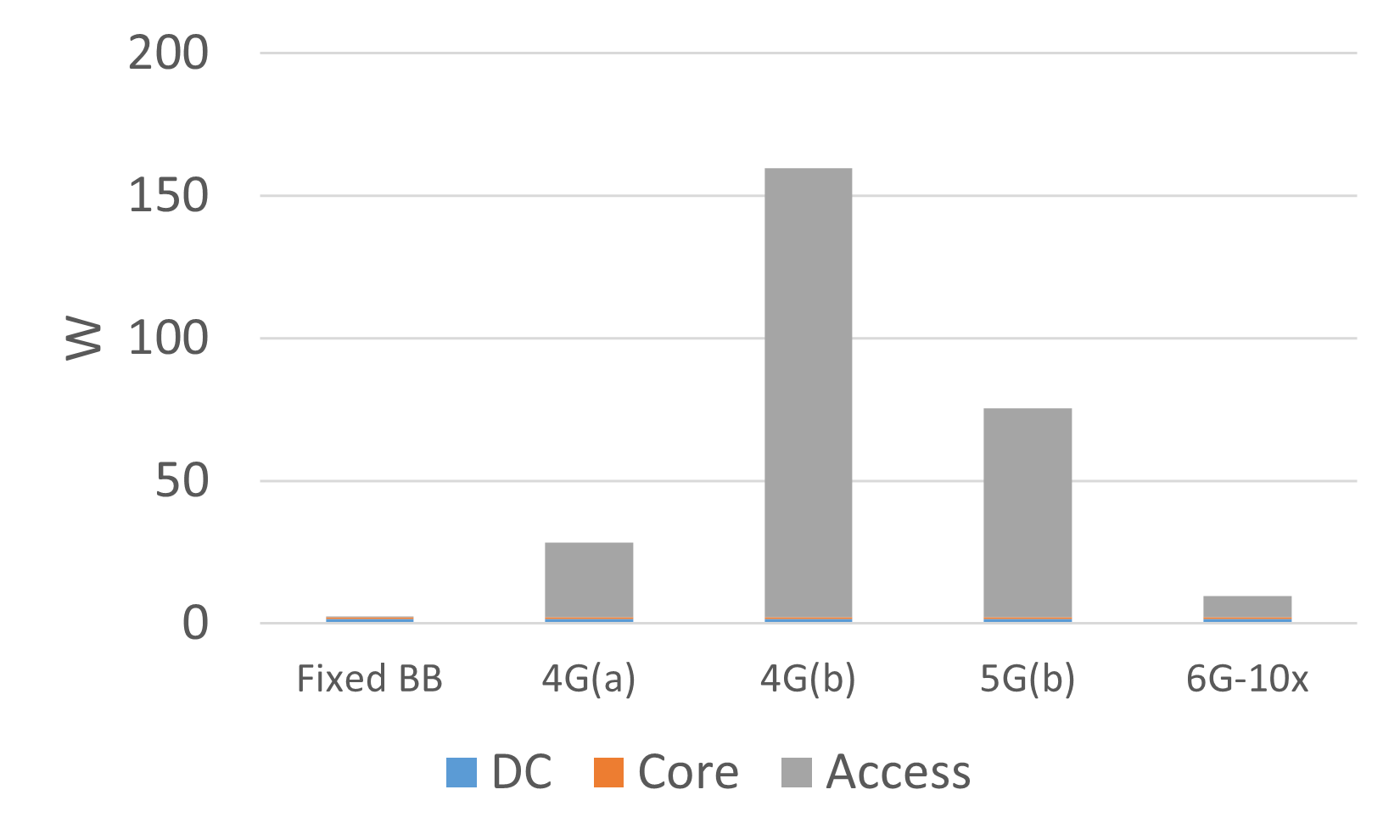}\\
{\footnotesize  \hspace{-0.22in}\small{(b)  4K to HD}}
\end{minipage}\
\end{tabular}
\end{center}
\caption[]{\protect Energy reduction from reducing video quality from 4K to FHD and SD, for different network access. For 4G/5G we consider the dense urban energy consumption data in Table~\ref{tab:energy_consumption}. We consider a small size  CDN.}
\label{fig:energy_reduction_different_access}
\end{figure}

Based on the energy consumption data in Table~\ref{tab:energy_consumption}, we can calculate the energy reduction  when the video quality is reduced from 4K to FHD and HD, assuming the following bitrates: $x_{\max,\text{4K} }=20$~Mbps, $x_{\max,\text{FHD} }=8$~Mbps, and $x_{\max,\text{HD} }=2.5$~Mbps.
The results are shown in Figure~\ref{fig:energy_reduction_different_access}(a) and~\ref{fig:energy_reduction_different_access}(b) for a reduction from 4K to FHD and from 4K to FHD, respectively.   
Our first  observation is that the energy reduction in the case of fixed broadband access is significantly smaller than for 4G and 5G access. 
Additionally, the reduction for 4G and 5G access is dominated by the energy reduction due to the mobile access network: the contribution is more than 98.5\% for 4G access and more than 97\% for 5G access, based on the  data from~\cite{Gol++23}. 
Even considering the significantly lower incremental consumption from ~\cite{Mal20},  cellular access contributes more than 92\% of the total energy reduction. Indeed, these results are for a small size CDN. The percentages would be higher for larger size CDNs.
This result implies that  \textit{for 4G/5G access,  incentives should focus on  mobile access energy consumption}.
Finally, Figure~\ref{fig:energy_reduction_different_access} shows that the energy reduction for 5G access is smaller than for 4G, which reflects the higher incremental energy efficiency of 5G technology.  

For 6G access, if we assume a 10x improvement compared to the 5G consumption shown in Table~\ref{tab:energy_consumption}, then  6G access contributes 25.1\% of the total energy consumption reduction for a small CDN size and 50.3\% for a very large CDN size. The contributions due to the energy consumption in the CDN are 56.7\% and 13.5\% for a small and very large CDN size, respectively. Hence, unlike the case of 4G and 5G, the significantly higher energy efficiency of 6G technology results in 6G access having a less dominant effect on the overall energy consumption and the  contribution of the CDNs becoming significantly larger. 
With a 100x increase of energy efficient 6G technology compared to 5G, the contribution of cellular access would be even smaller, while the CDN contribution would be more significant; In this case, an analysis similar to that in the next section for fixed broadband access and Section~\ref{sec:carbon_aware_dc_selection} for carbon-aware data center selection would be applicable. 
Nevertheless, it is important to note that the relative energy consumption  will depend on future improvements in energy efficiency of data centers.

\mynotex{Not sure if the projection for the energy consumption in 6G should be added to Table~\ref{tab:energy_consumption}.}

\subsection{Fixed broadband}

\label{sec:fixed_broadband}

\mynotex{
\begin{itemize}
\end{itemize}
}

\begin{figure}[tb]
    \centering
    \includegraphics[width=0.95\linewidth]{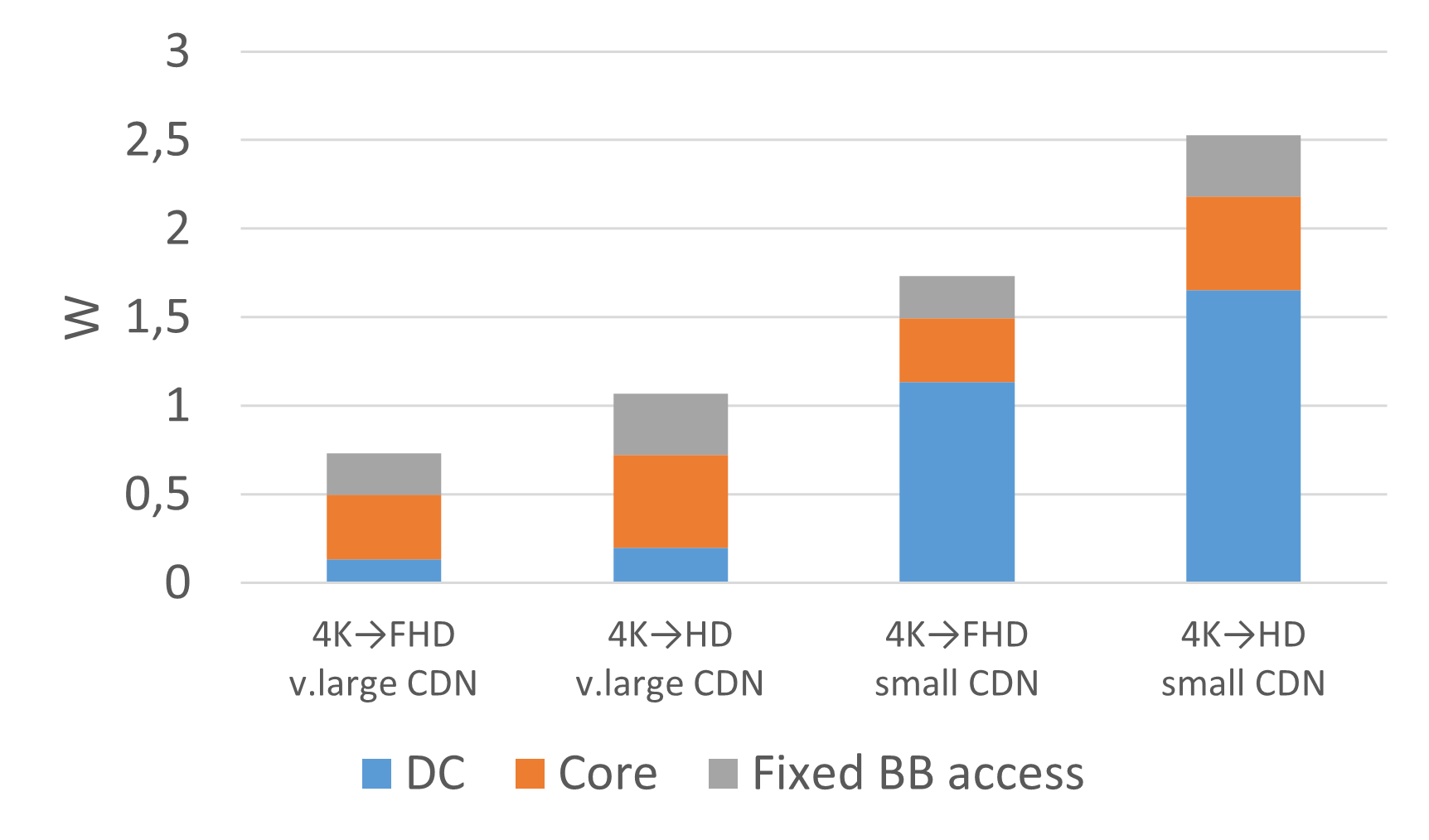}
    \caption{Fixed broadband access: energy reduction when video quality is reduced from 4K to FHD and to HD, for a small and very large CDN scenario. 
    }
    \label{fig:energy_reduction_fixed}
\end{figure}

The contribution of the different segments (fixed access, core, and data center) of an end-to-end path in the case of fixed broadband access are significantly different than 4G/5G access, Figure~\ref{fig:energy_reduction_fixed}.
This figure shows that a small CDN contributes to energy reduction by approximately 65\%, whereas the core network's contribution  is approximately 21\%.
On the other hand, the contribution of a very large CDN is approximately 18\%, while in this case the core network's contribution rises to 49\%.
These results suggest that \textit{for fixed broadband access, CDNs can have a significant contribution to the energy consumption reduction, whose magnitude depends on the size of the CDN}.  
Moreover, the above conclusion  motivates the investigation of scenarios where a video can be streamed from different data centers with different carbon intensities; this is investigated in the next section.

\section{Carbon-aware data center selection}
\label{sec:carbon_aware_dc_selection}

\mynotex{
\begin{itemize}




\item also note carbon intensity is time varying.

\item highlight conditions holds independent of access technology. however, for 4/5G access the contribution of the dc is very small.

\end{itemize}
}

The investigation in Section~\ref{sec:fixed_broadband} showed that the energy contribution of  CDNs can be significant for fixed broadband access, but the magnitude depends on the CDN size. Moreover,  prior work has suggested routing requests to data centers taking into account carbon emissions \cite{Zho++16,Elz++25}.
Based on this, the objective of this section is to present a simple model for carbon-aware selection of the data center that serves a video request.
The model captures the tradeoff between the lower carbon intensity of the remote data center compared to the local data center and the higher network cost from  the remote data center to the user.

\begin{figure}[tb]
\vspace{+0.02in}
    \centering
    \includegraphics[width=0.95\linewidth]{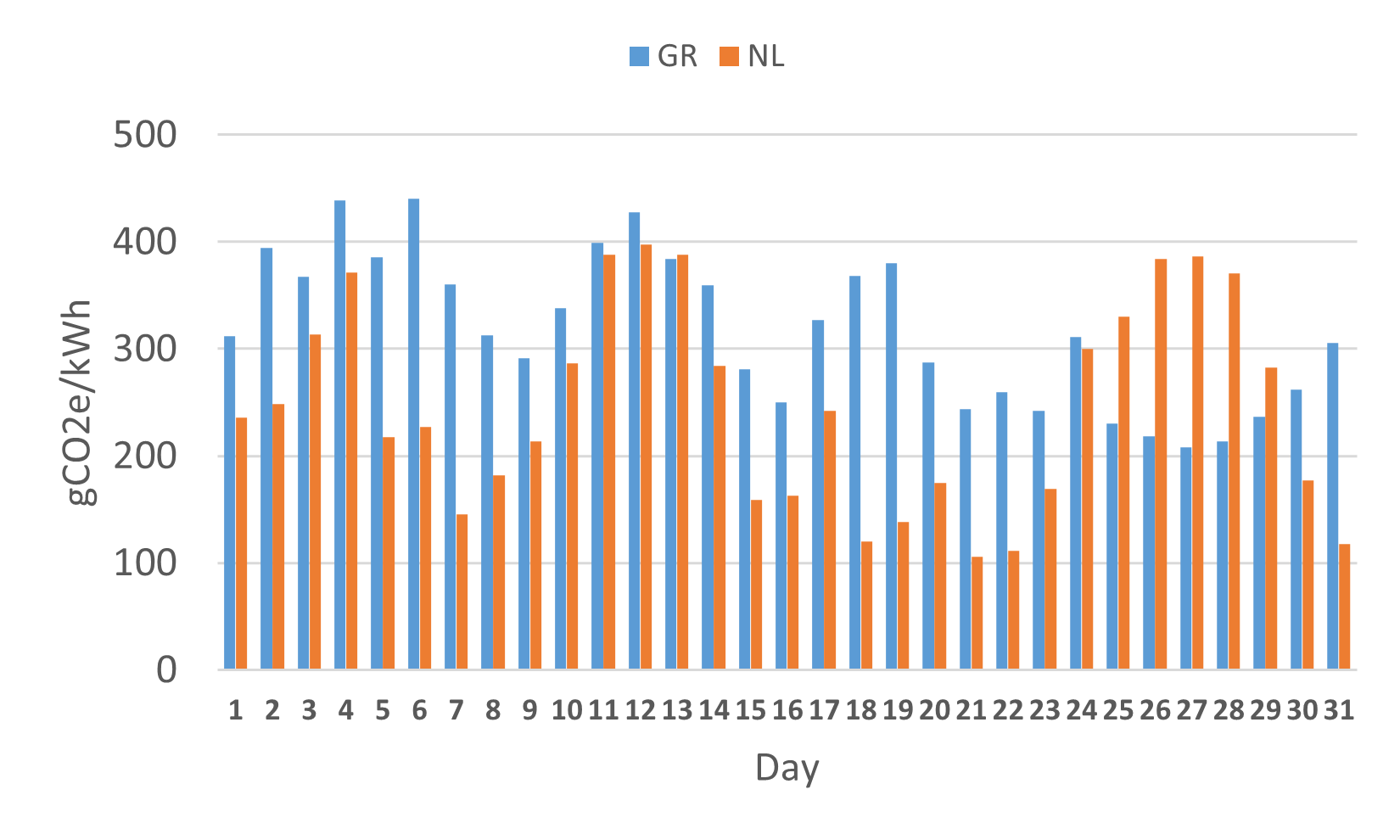}
     \vspace{-0.03in}
    \caption{Carbon Intensity in two countries. Data obtained from \texttt{electricitymaps.com}
    }
    \vspace{-0.03in}
    \label{fig:CI_day_GR_NL}
\end{figure}

We consider empirical measurements for the carbon intensity in two different countries: Greece and the Netherlands, Figure~\ref{fig:CI_day_GR_NL}, which show that there are days where the carbon intensity of energy production is lower in Greece, e.g., from the 25th to the 29th, and days where it is lower in the Netherlands, e.g., from the 1st to the 10th.

Assume a user residing in Greece requests a video. If the video request is served by a local data center in Greece with carbon intensity $CI_\text{l}$, then
the incremental carbon emissions to serve the video is 
\begin{equation}
(a+c+d_{\text{l}}) \cdot CI_{\text{l}} \, ,
\label{eq:local}
\end{equation}
where $a$, $c$, and $d_{\text{l}}$ are the incremental energy consumption of the access network, the core network, and the local data center. 
The model holds for any access technology, but we consider fixed broadband access, since for this case the energy consumption at the data center can be a significant part of the end-to-end energy consumption, as discussed in Section~\ref{sec:fixed_broadband}. 

An alternative  to serving the  video from a local data center located in the same country as the user is to serve the video from a data center located in a different country (the Netherlands in our experimental investigation), in which case the incremental energy consumption is
$$
(a+c) \cdot CI_{\text{l}}+(c+d_\text{r}) \cdot CI_{\text{r}} \, .
$$
In the last equation, carbon emissions due to the user’s access network and part of the core network transmission depend on the carbon intensity at the user's geographic location. However,  the carbon emissions due to the  part of the core network at the data center side and  the incremental energy consumption at the data center's CDN depend on the carbon intensity at the data center's geographic location. Compared to (\ref{eq:local}), we assume that the core energy consumption is now double, due to the longer network distance from the data center. The model can be extended with a more detailed attribution of the core network costs and to consider that the interconnection involves multiple intermediate countries with possibly different carbon intensities; nevertheless, the simple model sufficiently captures the tradeoff of serving a video from a remote data center with a smaller carbon intensity while incurring  higher core network energy consumption due to the longer distance from the data center.

From the above two equations, the carbon emissions when a remote data center serves the video request are lower than the emissions when a local data center serves the video if the following holds:
%
%
\begin{align}
(a+c) \cdot CI_{\text{l}}+(c+d_{\text{r}}) \cdot CI_{\text{r}} & <
(a+c+d_{\text{l}}) \cdot CI_{\text{l}}    \Rightarrow \nonumber \\
(c+d_{\text{r}}) \cdot CI_{\text{r}} & < d_{\text{l}} \cdot CI_{\text{l}}
\Rightarrow \nonumber \\
\frac{CI_{\text{r}}}{CI_{\text{l}}} & < \frac{d_{\text{l}}}{c+d_{\text{r}}} \label{eq:dc_selection}
\end{align}
%
Note that the last condition depends on the carbon intensity and energy consumption at the local and remote data center, and on the energy consumption in the core network, due to the higher network costs from the remote data center; however, note that the condition is  independent of the energy consumption at the access network, since its contribution is the same if the video is streamed from the local or from the remote data center.

\begin{figure}[tb]
\vspace{+0.02in}
    \centering
    \includegraphics[width=0.95\linewidth]{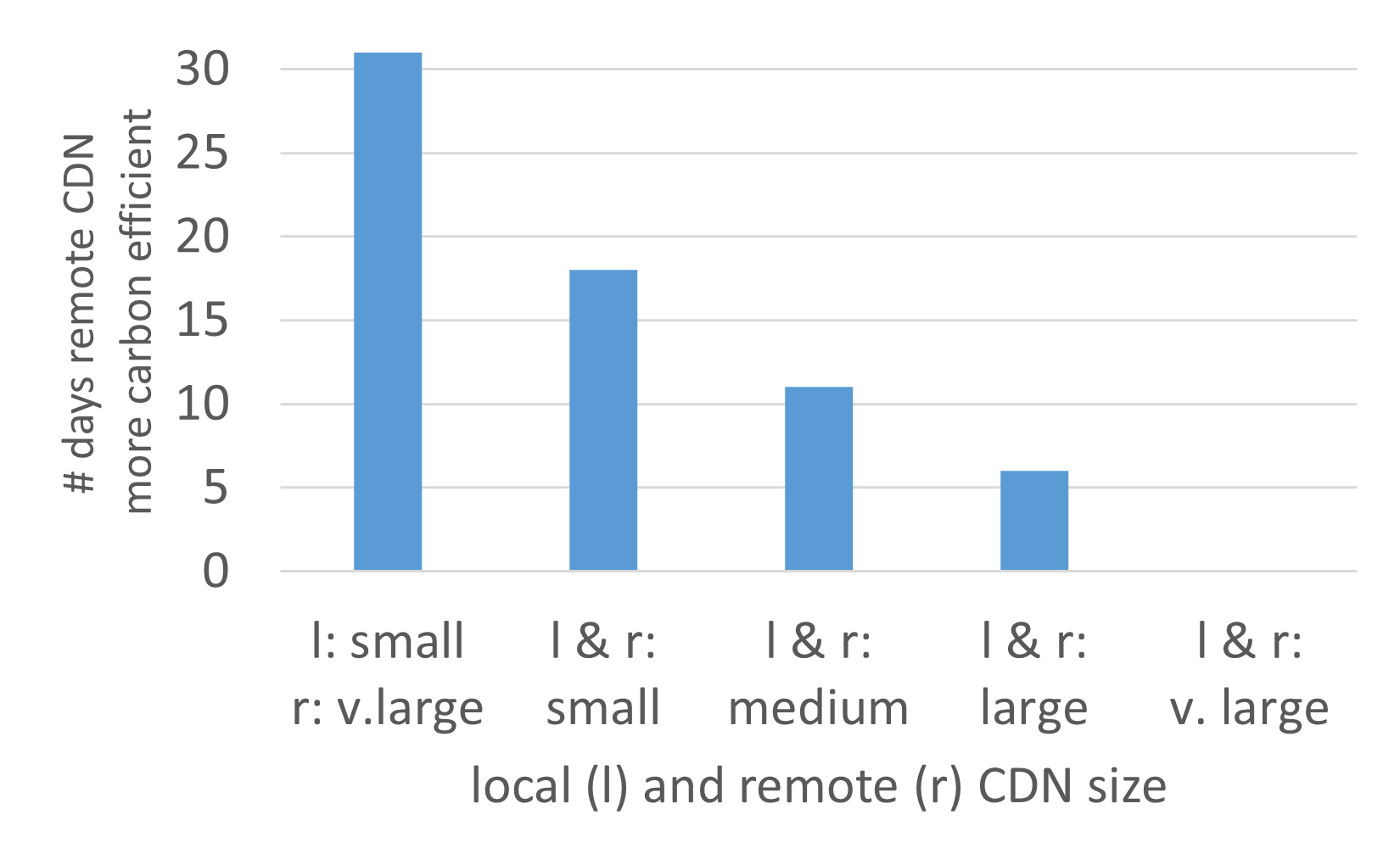}
    \caption{Days when the remote CDN selection is more carbon efficient. 
    }
    \label{fig:days_remote_cdn}
\end{figure}

Using  condition (\ref{eq:dc_selection})  and the empirical measurements shown in Figure~\ref{fig:CI_day_GR_NL}, we can determine, for each day of the month, whether the carbon emissions are lower when the video is served by the local data center or by the remote data center. The number of days during which the remote data center achieves  lower carbon emissions is  shown in Figure~\ref{fig:days_remote_cdn}. 
When both the local and the remote CDNs are very large, selecting the local data center yields lower carbon emissions for all days of the month. This occurs because the energy consumption in both CDNs is low, and the higher data transmission cost from the remote CDN makes it less preferable. 
At the other extreme, when the local CDN is small and the remote CDN is very large, the benefits of reduced energy consumption at the remote CDN outweigh the higher core network cost, while also benefiting from the lower carbon intensity at the remote CDN for some days of the month. For intermediate CDN sizes (small, medium, and large), selecting the remote CDN provides lower carbon emissions for  days ranging from 6 (when both the local and the remote CDN are large) to  18 (when both the local and the remote CDN are small); this occurs for the following reason:  when the energy consumption at the CDN is high,  which is the case with small CDNs, the benefits of lower carbon intensity at the remote CDN are higher, hence selecting the remote CDN is more carbon efficient for more days of the month.

\section{Energy reduction and incentives  for different user types}
\label{sec:incentives}

\mynotex{
\begin{itemize}
\item amongst the goals of this subsection is to show how carbon-aware dc selection impacts the incentives.
\item modify subsection titles: first subsection applies to 4G/5G but also to case of very larger local and remote CDN, Figure~\ref{fig:days_remote_cdn}. DONE
\end{itemize}
}

In this section we present an approach for determining the incentives based on the daily variations of the carbon intensity reducing the video quality,  how these incentives  depend on the type of users, namely high-quality and green (environmental conscious), and the resulting bitrate reduction. Although our analysis considers the carbon intensity in the timescales of a day, the approach can be applied to different timescales. 

\subsection{Local CDN only}
\label{sec:incentives_local_CDN_only}

\mynotex{
\begin{itemize}
\item we apply method to only one month. Should we add one result for a different month? Space might be a constraint.
\end{itemize}
}

We first consider the case where the video request is served from a local data center located in the same country where the user is located. This can be the case of 4G/5G access, when the energy consumption along the end-to-end path is dominated by the cellular access network, hence considering remote data center alternatives does  not practically impact the energy consumption and carbon emissions.
We consider the  empirical carbon intensity measurements  for  Greece in  December 2024,  Figure~\ref{fig:CI_day_GR_NL}. 

\begin{figure}[tb]
\begin{center}
\begin{tabular}{c}
\begin{minipage}[b]{1\linewidth}
\centering
\hspace{-0.28in}
\includegraphics[width=2.9in]{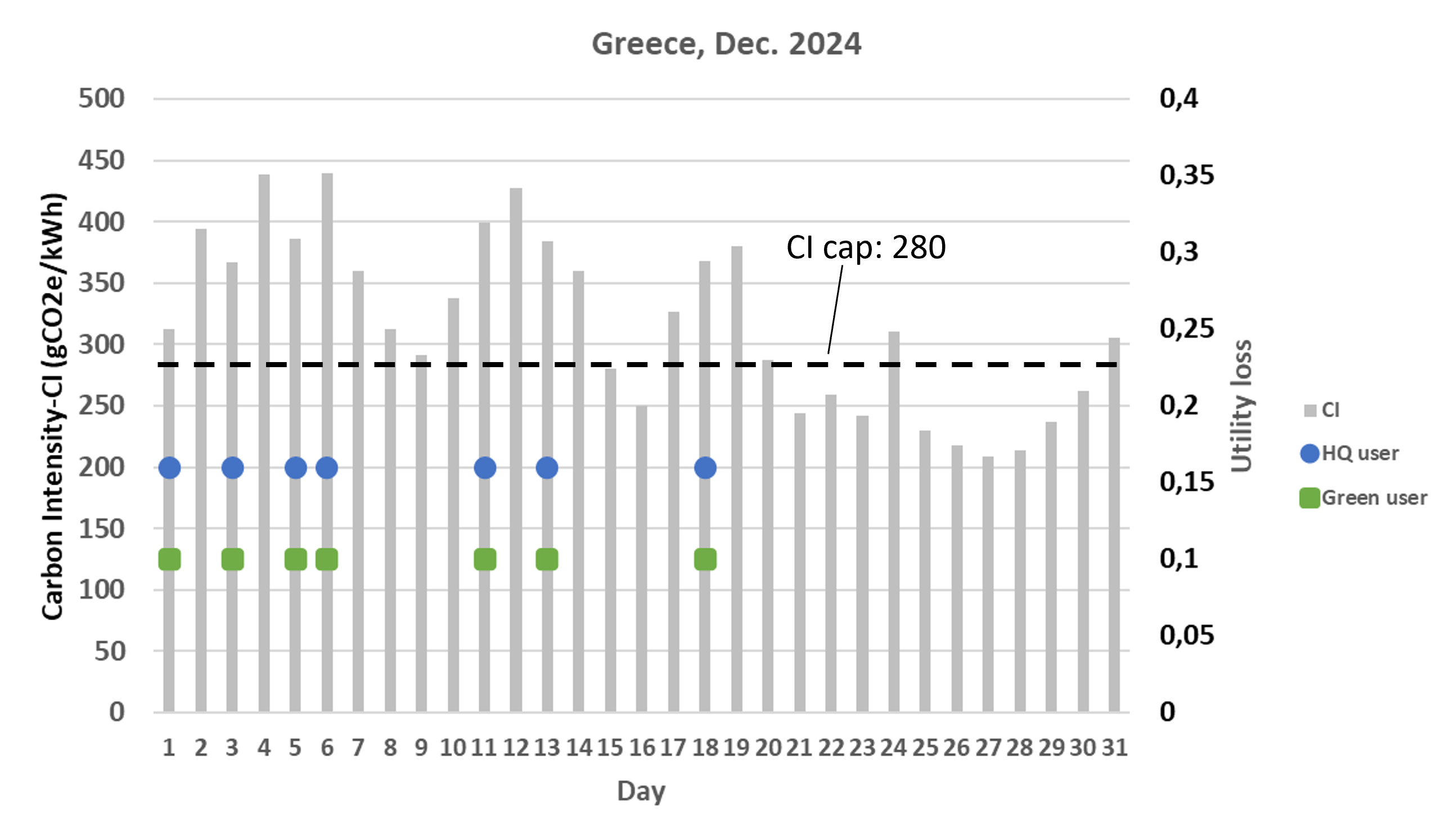}\\
{\footnotesize  \hspace{-0.22in}\small{(a) incentive strategy 4K$\rightarrow$FHD }}
\end{minipage} \\ \\
\begin{minipage}[b]{1\linewidth}
\centering
\hspace{-0.28in}
\includegraphics[width=2.9in]{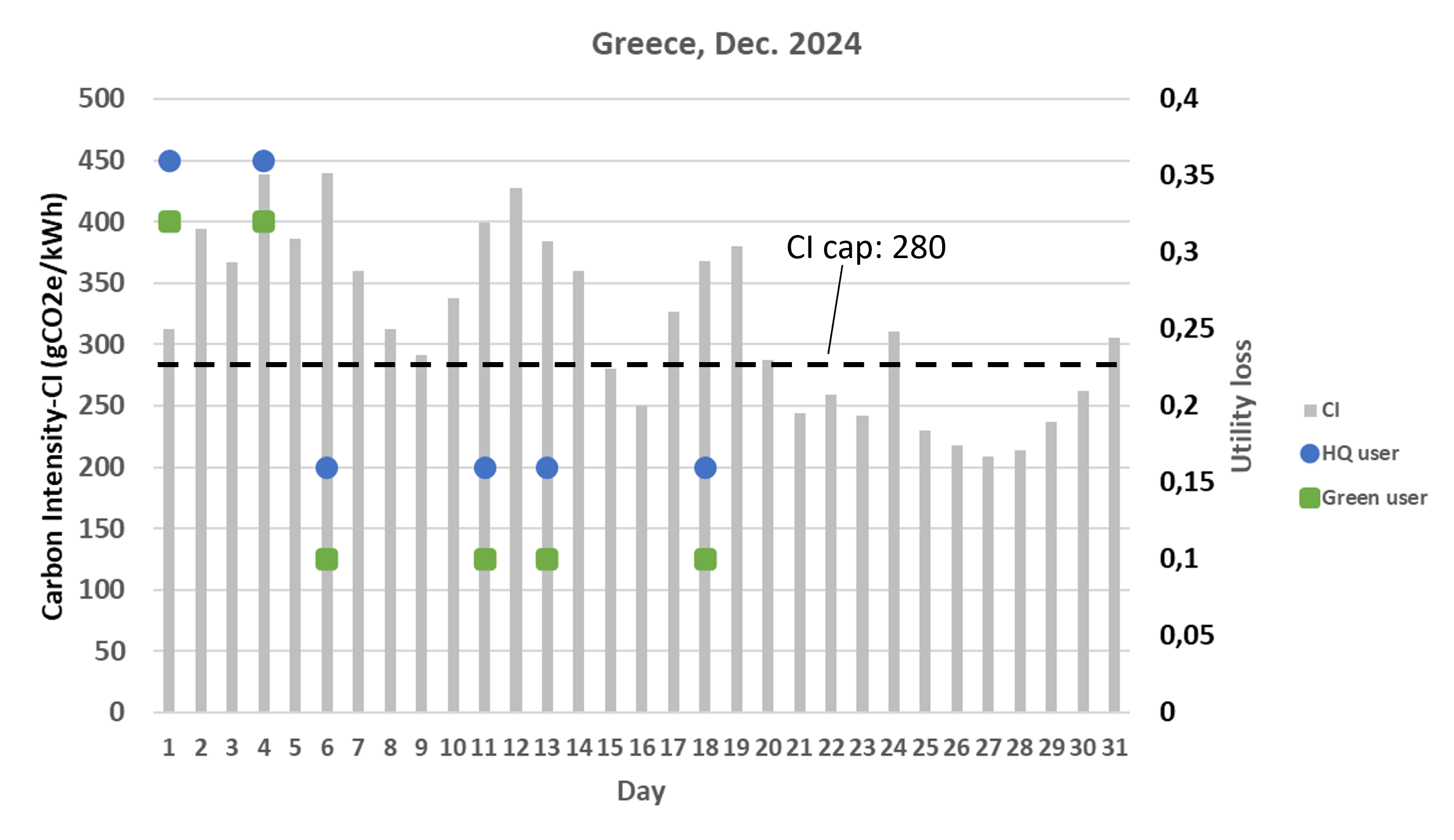}\\
{\footnotesize  \hspace{-0.22in}\small{(b) incentive strategy 4K$\rightarrow$FHD,HD }}
\end{minipage}
\end{tabular}
\end{center}
\caption[]{\protect Utility loss and incentives  for one location and two incentive strategies. The incentives have the goal to maintain the average carbon intensity  below 280 gCO2e/kWh.}
\label{fig:utility_loss_day}
\end{figure}

Consider that the target is to maintain the average carbon intensity below the threshold 280 gCO2e/kWh. 
This carbon cap can be influenced  from  some public authority, such as through  carbon certificates for emissions reduction.
For simplicity, we assume that video requests are uniformly distributed across the days of the month. 
To determine the days that the video quality needs to be reduced we follow a sequential procedure that starts from the first day and updates a carbon intensity budget by adding the intensity surplus when the carbon intensity is below the cap and subtracting the excess amount when the carbon intensity is above the cap. If the budget would be negative for a particular day, then for that day the video quality must be reduced  to maintain a positive carbon intensity budget. We consider two strategies for reducing the video quality: The first strategy reduces the video quality from 4K to FHD only. The second strategy reduces the video quality from 4K to HD for the first two days and from 4K to FHD for the remaining days when a reduction is necessary; note that the specific selection of days does not affect our conclusions, as we discuss below. 
Figure~\ref{fig:utility_loss_day}(a) shows the days when the video quality needs to be reduced according to the first strategy, referred to as 4K$\rightarrow$FHD. Note that the days of reduced video quality are independent of the access technology and the energy consumption along the end-to-end path from the data center to the user, and depend solely on the carbon intensity values across the different days and the carbon intensity cap.

Table~\ref{tab:two_strategies} shows that the first strategy requires the video quality to be reduced to FHD on seven days of the month. Note that the days at which the video quality must be  reduced  are independent of the user type.
However, the user type determines the utility loss from the reduction: 1.12 for a high-quality user and 0.7 for a green user, which corresponds to an average reduction of the utility approximately 3.6\% and 2.2\% over the month, for high-quality and green users, respectively.
\begin{table}[b]
\begin{center}
\caption{Utility loss and bitrate reduction for the two incentive strategies}
\label{tab:two_strategies}      
\begin{tabular}{|l|c|c|l|}
 \hline 
 Incentive  & \# days   & bitrate  & Total utility loss       \\ 
 strategy &  & reduction &  (avg monthly reduction)       \\ 
 \hline \hline
4K$\rightarrow$FHD  & 7 & 13.5\% & HQ user: 1.12 (3.6\%) \\
             &  &  & Green user: 0.70 (2.2\%) \\ 
             \hline
4K$\rightarrow$FHD,HD  & 6 & 13.4\% & HQ user: 1.36 (4.4\%) \\
            &  &  & Green user: 1.04 (3.4\%) \\ 
             \hline 
\end{tabular}
\end{center} 
\end{table}

Figure~\ref{fig:utility_loss_day}(b) shows the days that the video quality needs to be reduced with the second strategy, which decreases the video quality from 4K to HD (instead of FHD) for the first two days and to  FHD for the remaining days. A higher video quality reduction in the first two days of decreased quality gives a higher  carbon intensity surplus, which  results in the video quality being reduced in six days,  one less than the number of days  the quality must be reduced with the first strategy, Table~\ref{tab:two_strategies}. The average bitrate reduction percentage for the whole month is approximately equal for the two strategies, hence the energy and carbon emissions reduction is also approximately equal.
However, the second strategy incurs significantly higher utility loss for both the high-quality and the green users:  4.4\% average utility loss (second strategy) compared to  3.6\%  (first strategy) for high-quality users and  3.4\% (second strategy) compared to 2.2\% (first strategy) for green users.
From the above, the incentives that must be offered to users with the second strategy is higher than the first strategy, while achieving approximately the same bitrate reduction. This makes the first strategy preferable for the provider. The above result is due to the shape of the utility curve in Figure~\ref{fig:utility}, whose slope is higher for lower bitrates. As a result, the total utility loss, which determines the incentives that must be offered, is higher for any scheme that includes a reduction of the video quality to HD, compared to the first strategy that reduces the video quality to FHD, despite achieving approximately the same bitrate reduction. 
%


The impact of different intensity caps is shown in Table~\ref{tab:caps}. As expected, a smaller average carbon intensity cap results in more days in which the video quality must be reduced and a higher total utility loss, which must be compensated with equal incentives.
In return, a smaller average carbon intensity cap yields a higher bitrate reduction, which provides a higher (proportional) reduction of the incremental energy consumption and carbon emissions.
For a very low average carbon intensity cap, it may happen that the video quality must be  reduced for all days of the month.
In this degenerative case of an even lower carbon intensity cap, a reduction to even lower video qualities, namely HD, would be necessary.
\begin{table}[tb]
\begin{center}
\caption{Utility loss and bitrate reduction for different carbon intensity caps}
\label{tab:caps}
\begin{tabular}{|l|c|c|l|}
 \hline 
 CI cap  & \# days   & bitrate  & Total utility loss       \\ 
  &  & reduction &  (avg monthly reduction)       \\ 
 \hline \hline
280 & 7 & 13.5\% & HQ user: 1.12 (3.6\%) \\
             &  & & Green user: 0.70 (2.2\%) \\  \hline
240 & 12 & 23.2\% & HQ user: 1.92 (6.2\%) \\
             &  &  & Green user: 1.20 (3.9\%) \\  \hline
200 & 18 & 34.8\% & HQ user: 2.88 (9.3\%) \\
             &  &  & Green user: 1.80 (5.8\%) \\  \hline
160 & 25 & 48.4\% & HQ user: 4.00 (12.9\%) \\
             & &  & Green user: 2.50 (8.1\%) \\  \hline
\end{tabular}
\end{center}
\end{table}

\subsection{Local and remote CDN}
\label{sec:incentives_fixed_broadband}
\label{sec:incentives_local_and_remote_CDN}

\begin{figure}[tb]
\begin{center}
\begin{tabular}{c}
\begin{minipage}[b]{1\linewidth}
\centering
\hspace{-0.28in}
\includegraphics[width=2.9in]{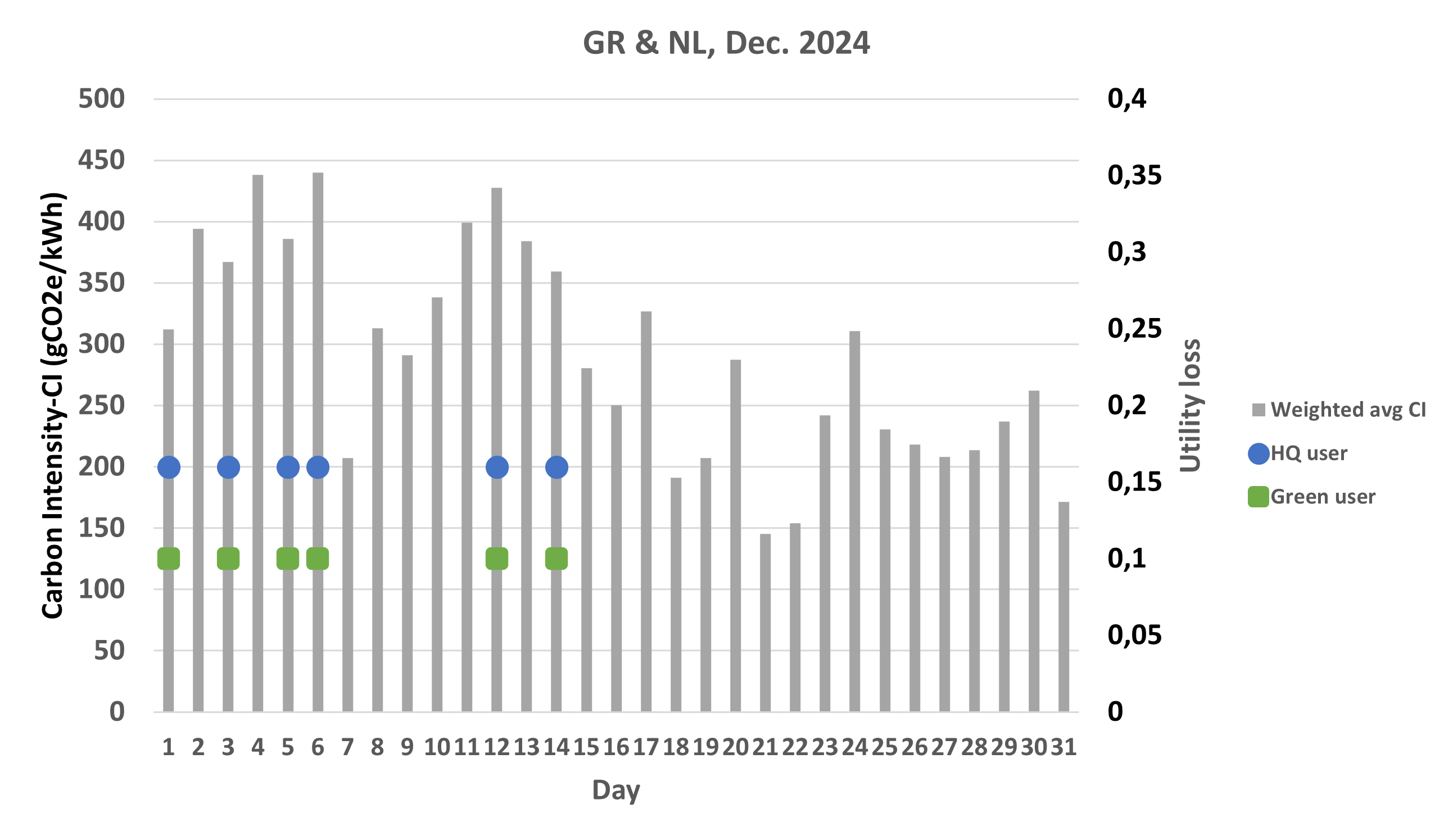}\\
{\footnotesize  \hspace{-0.22in}\small{(a) Local, remote: large CDN }}
\end{minipage} \\ \\
\begin{minipage}[b]{1\linewidth}
\centering
\hspace{-0.28in}
\includegraphics[width=2.9in] {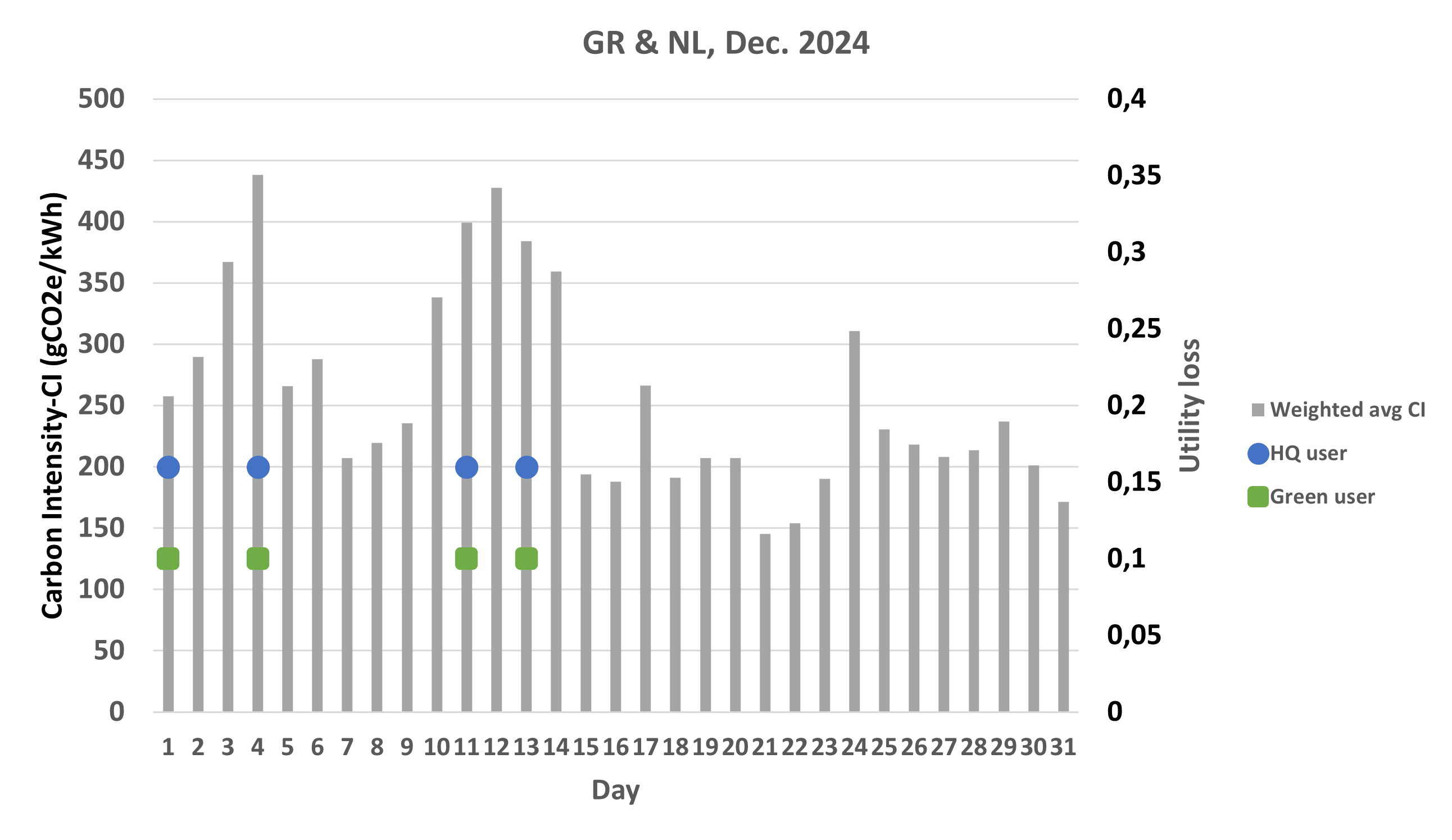}\\
{\footnotesize \hspace{-0.22in}\small{(b)  Local, remote: small CDN}}
\end{minipage} \\ \\
\begin{minipage}[b]{1\linewidth}
\centering
\hspace{-0.28in}
\includegraphics[width=2.9in]{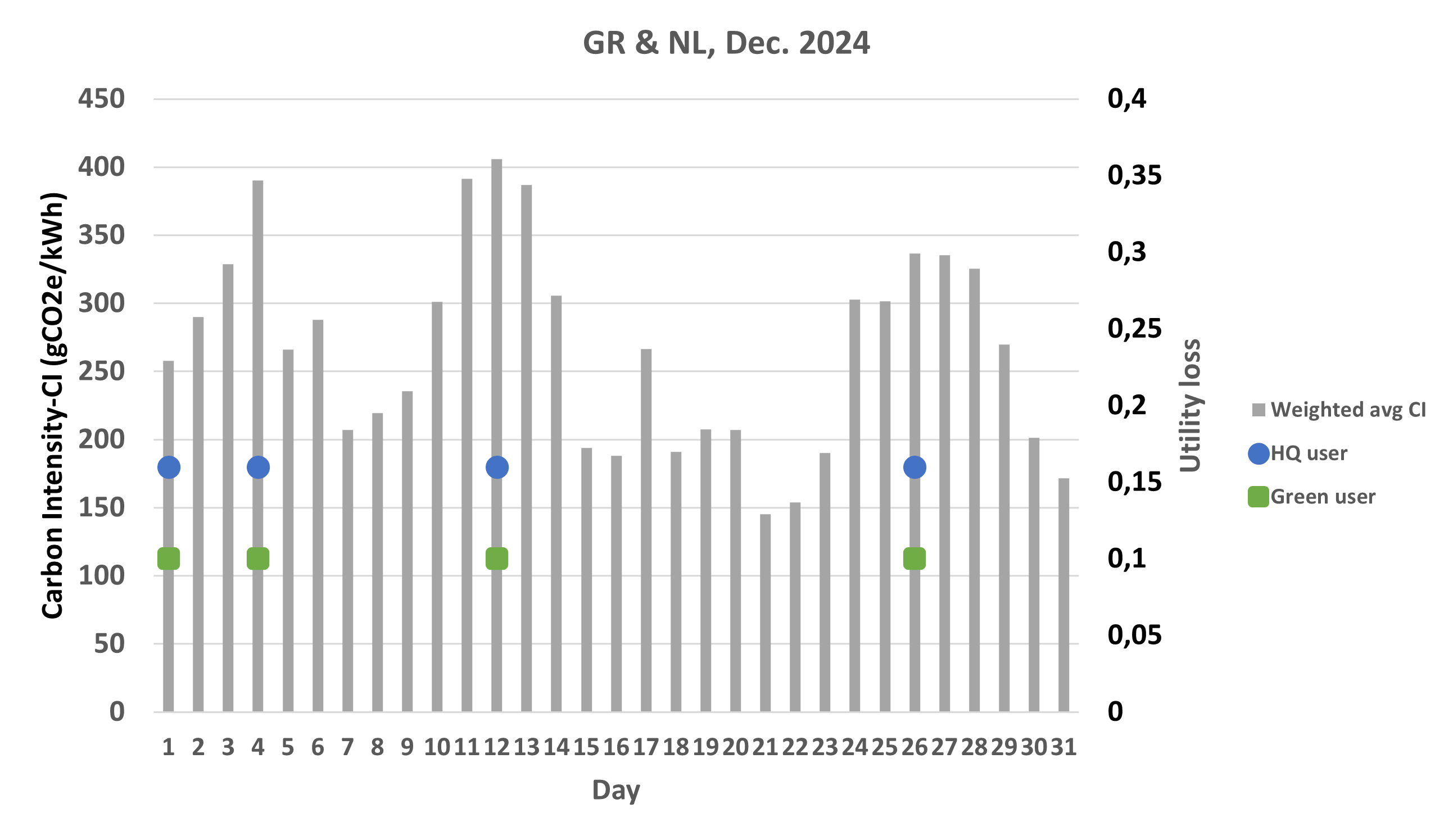}\\
{\footnotesize  \hspace{-0.22in}\small{(c)  Local: small CDN, remote: very large CDN}}
\end{minipage}
\end{tabular}
\end{center}
\caption[]{\protect Utility loss and incentives for different local and remote CDN sizes.}
\label{fig:utility_loss_day_cdn}
\end{figure}

Next we consider the case where a video can be streamed by either a local or remote CDN, which are located in data centers with different carbon intensities.
We consider the following combinations of CDN sizes: i) both the local and remote CDN are large, ii) both are  small, and iii) the local CDN is small and the remote is large, Figure~\ref{fig:utility_loss_day_cdn}.
As in the previous subsection, we consider  the target of maintaining an average  carbon intensity below  280 gCO2e/kWh. 
Based on the results of the previous subsection, we consider only the incentive strategy where the video quality is reduced from 4K to FHD. 
Note that unlike the case where a video is streamed from a local CDN, 
when two CDNs are considered the days that the video quality must be reduced to satisfy a carbon intensity cap depend on the energy consumption at the local and remote CDNs, the energy consumption at the core network, and the relative carbon intensity at the local and remote data center where the CDNs reside; these factors appear in 
condition (\ref{eq:dc_selection}) that determines whether the local or remote CDN provides lower carbon emissions for streaming a video.

Comparison of Figure~\ref{fig:utility_loss_day_cdn}(a) with Figure~\ref{fig:utility_loss_day}(a)  shows the impact of exploiting a remote CDN with lower carbon intensity when both the local and remote CDNs are large. For large CDNs, the remote CDN is more energy and carbon efficient in 6 out of the 31 days of the month, Table~\ref{fig:days_remote_cdn}. The flexibility of being able to select a remote CDN reduces the number of days that the video quality must be reduced by one, compared to the case where only a local  CDN is available. When both the local and remote CDNs are small the impact is larger: The video quality needs to be reduced in four days, Figure~\ref{fig:utility_loss_day_cdn}(b), compared to seven days in the case of a single CDN, Figure~\ref{fig:utility_loss_day}(a).

Finally, when the local CDN is small and the remote is very large, Figure~\ref{fig:utility_loss_day_cdn}(c), we see that the number of days in which the video quality needs to be reduced is four, the same as when both CDNs are large. This shows that the improvements (in terms of decreasing the number of days the video quality must be reduced) reach a saturation point when the remote CDN is used for more than 18 days.
Figure~\ref{fig:utility_loss_day_cdn}(c) also shows that the simple  condition for choosing whether to select a local or remote CDN given by (\ref{eq:dc_selection}) captures the following tradeoff: Although the carbon intensity in Greece is smaller  for some days of the month, which include the period from 25th to 29th, Figure~\ref{fig:CI_day_GR_NL}, it is still preferable to use the remote data center in the Netherlands since the very large size of its CDN  provides smaller energy consumption that outweighs its higher carbon intensity during those days.

\mynotex{
Fig 8(c), local small CDN and remote very large CDN captures an interesting tradeoff: although the carbon intensity in Greece is smaller than the Netherlands for some days of the month, which include 25th to 29th, Figure~\ref{fig:CI_day_GR_NL}, it is nevertheless preferable to use the CDN in the Netherlands since its very large size of the CDN  provides smaller energy consumption which out-ways the higher carbon intensity for those days.  
}

\section{2-tier subscription model with carbon-aware discounts}
\label{sec:2-tier}

Our discussion up to now has focused on when video quality reduction incentives should be offered and how much the incentives should be. For the first, incentives should be offered based on the carbon intensity and the average carbon intensity cap. For  the second,  incentives  should compensate users for the utility loss incurred due to the reduced  video quality. 
In this section we describe a 2-tier approach for offering users incentives in the form of subscription options with discounts and associated rewards in exchange for lowering the quality of up to some maximum percentage of videos streamed within a time period, e.g., one month. Such subscriptions give the  provider the flexibility to stream videos at lower quality during periods (days) when the carbon intensity is high, in order to reduce their carbon emissions and satisfy an average carbon intensity cap.
An advantage of this indirect form of incentives is that it does not require actions from users on a per-video basis,  which have a high accounting overhead and lead to user fatigue that can reduce the impact of incentives. Moreover, note that our proposed approach involves only two tiers: a high-quality tier and a low quality tier where the video resolution is directly lower than the highest resolution.
This is justified by the results in Section\ref{sec:incentives_local_CDN_only} and Table~\ref{tab:two_strategies}, where we showed that a reduction of the video quality below the  resolution  directly lower than the highest resolution that achieves the maximum user satisfaction incurs a higher utility loss, hence the provider needs to provide more incentives, without any additional gains.

Figure~\ref{fig:utility_loss_day}(a) shows the incentives that must be offered for users to reduce their video quality  to achieve a cap on the carbon intensity equal to 280 gCO2/kWh. A direct  approach would be to offer users the incentives on a per-video basis. For example,  the provider can present, in the form of a notification on the video client,  an incentive to reduce the video quality on  days when the carbon intensity exceeds the cap.
Instead of such per-video incentives, we assume that the provider has a forecast for the carbon intensity in the one month period. We also assume that video requests are uniformly distributed across all days of the month, although the analysis below can be adapted to an uneven distribution of videos in one month.
Figure~\ref{fig:utility_loss_day}(a) indicates that in  7 out of the 31 days, which gives  a percentage approximately 22.5\%, 
the video quality must be reduced to FHD. Hence, the provider can offer users a subscription option that allows the provider to reduce the quality of up to 22.5\% of  videos to FHD in  one month. Note that  22.5\%  is an upper bound and the provider can eventually reduce the  quality for a smaller percentage of videos, when  the actual carbon intensity is lower than the forecast. Moreover, note that the percentage of videos with reduced quality, when all videos are streamed from a local data center, depends solely on the carbon intensity and the average carbon intensity cap, hence are independent of the access technology and  the incremental energy consumption along the end-to-end path from the data center to the user.

The provider must offer appropriate incentives to encourage users to select the above 2-tier subscription model. These incentives should be at least the total utility loss that is incurred by users due to the reduced video quality for the 6 days in Figure\ref{fig:utility_loss_day}(a), which are shown in Table~\ref{tab:two_strategies}: 1.12 for high-quality users and 0.7 for green users. Hence, the incentives should be lower for green users compared to high-quality users, which as noted in Section\ref{sec:incentives} illustrates the importance of educating users on  sustainability and environmental consciousness.
The incentives can include a discount on the 2-tier subscription option, relative to the subscription option where all videos are delivered with 4K quality. This discount corresponds to the variable $s_{\delta \text{quality}}$ in (\ref{eq:net_utility_loss}).
The discount can be applied to the part of the price of the subscription that streams all videos at 4K, and can be proportional to the average bitrate reduction for the whole month, which as shown in Table~\ref{tab:two_strategies} is 13.5\%, since this reflects the reduced energy consumption costs from reducing the video quality. However, note that the overall discount can be much smaller than 13.5\%, due to the idle energy consumption of devices: This can be significant, higher than 80-90\% for wired and cellular network devices~\cite{Mal20,Gol++23}, but are lower, 35\% and decreasing towards 27\%, for data centers/CDNs~\cite{She++24}.

In addition to the above discount, which corresponds to $s_{\delta \text{quality}}$ in (\ref{eq:net_utility_loss}), the provider must offer incentives to compensate for the remaining amount of the utility loss, which is equal to the net utility loss given by (\ref{eq:net_utility_loss}).
The net utility loss can be provided to users in the form of carbon rewards or as an additional monetary discount on top of the 13.5\% discount due to the average bitrate reduction. The carbon rewards can be exchanged for monetary discounts in other services, tickets in carbon emission lotteries, etc.

The above analysis was for a  specific month, December 2024, which corresponds to the carbon utility loss and incentives shown in Figure\ref{fig:utility_loss_day}(a). The provider can  design a 2-tier subscription model that refers to multiple months. This would require the application of the methodology in Section~\ref{sec:incentives_local_CDN_only} to the forecast over a period of multiple months, followed by the aggregation of the incentive results, namely the percentage of videos with reduced quality and the total utility loss, over the multi-month period.

The discussion up to now considered the case where only a  local CDN is used to stream videos. If CDNs in multiple data center locations are considered, then the methodology in Section~\ref{sec:incentives_local_and_remote_CDN} can be followed. Recall that in this case an estimate of the incremental energy consumption of the local and remote CDNs must  along with the core network energy consumption is necessary, in order to apply the CDN selection condition given by (\ref{eq:dc_selection}). Unlike the local only CDN case, the percentage of  videos that must be streamed at lower quality depends on these parameters, in addition to the carbon intensity and the average carbon intensity cap.

A user typically has a subscription with a video service provider. Such subscriptions can take the form of the 2-tier subscription format discussed in this section.
However, as investigated in Section~\ref{sec:energy_consumption_e2e} the energy consumption reduction can be due to any segment along the end-to-end video delivery path.
A service provider can obtain measurements from the operators of intermediate segments of the end-to-end path by using mechanisms currently being investigated in 3GPP~\cite{Sul25} and the IETF~\cite{Rod++25}.

\section{Conclusions}
\label{sec:conclusions}

\mynotex{
\begin{itemize}
\item It is preferable to provide contracts that offer reduced video quality at a resolution that is immediately lower than the highest resolution (that achieves the maximum utility or user satisfaction).
\item For the 2-tier subscription that provides incentives for reduced video quality, when the video is streamed from a local data center (with a carbon intensity the same as other segments of the end-to-end path to the user),  the maximum percentage of videos streamed at a lower quality depend solely on the carbon intensity throughout a period (month) and the cap on the average carbon intensity over that period.
\item The total incentives that must be provided to users selecting the carbon-aware 2-tier subscription, in addition to the carbon intensity variation and the corresponding cap of the average carbon intensity, also on the user valuation for video quality, hence the type of user: it is larger for high-quality users compared to green (environmental conscious) users.  
\item When a video can be streamed from  data centers with  different carbon intensity,  the maximum percentage of videos streamed at a lower quality and the incentives also depend on the relative carbon intensities and energy consumption at the data centers and the higher energy consumption due to the longer network path to remote data centers. 
\end{itemize}
}

We considered  data  for the energy consumption along the end-to-end video delivery path  and discussed its implications on the design of incentives to balance user QoE and carbon emission levels.
%
Based on the proposed incentive model, we presented a method for  designing practical 2-tier subscriptions that give providers the flexibility to serve up to some percentage of videos  at a lower quality during periods of high carbon intensity, in exchange for providing users a subscription discount  and carbon rewards that balance their utility loss incurred from viewing videos at a lower quality.

\balance
Ongoing work is investigating how the monetary discount and  carbon rewards presented to users can be shared by network operators, CDNs, and streaming providers whose resources are used for video delivery.  
We are also investigating how the specific form of rewards, such as environmental points that can be exchanged for discounts or services and  carbon-based lottery tickets, can impact  user behavior. 

\mynotex{Add future work}

\mynotex{
Our key findings include the following:
\begin{itemize}
\item it is preferable to provide contracts that offer reduced video quality at a resolution that is immediately lower than the highest resolution (that achieves the maximum utility or user satisfaction).
\item for the 2-tier subscription that provides incentives for reduced video quality, when the video is streamed from a local data center (with a carbon intensity the same as other segments of the end-to-end path to the user),  the maximum percentage of videos streamed at a lower quality depend solely on the carbon intensity throughout a period (month) and the cap on the average carbon intensity over that period.
\item The total incentives that must be provided to users selecting the carbon-aware 2-tier subscription, in addition to the carbon intensity variation and the corresponding cap of the average carbon intensity, also on the user valuation for video quality, hence the type of user: it is larger for high-quality users compared to green (environmental conscious) users.  
\end{itemize}
}

\section*{Acknowledgment}

This work has been partly developed in the scope of the project EXIGENCE, which has received funding from the Smart Networks and Services Joint Undertaking (SNS JU) under the European Union (EU) Horizon Europe research and innovation programme under Grant Agreement No 101139120. Views and opinions expressed are however those of the author(s) only and do not necessarily reflect those of the EU or SNS JU.

\end{document}